\documentclass{eptcs}
\providecommand{\event}{ITRS 2014}
\usepackage{bussproofs}
\usepackage[cmex10]{amsmath}
\usepackage{amsthm}
\usepackage[utf8]{inputenc}
\usepackage{amssymb,wasysym,stmaryrd}
\usepackage{rotating}
\usepackage[T1]{fontenc}
\usepackage{tikz-qtree}
\usepackage[justification=centering,labelfont=bf]{caption}

\newcommand{\G}{$\mathcal{G}\,=\,\langle \Sigma, \mathcal{N},\mathcal{R},S \rangle\ $}
\newcommand{\collapse}[1]{\vert #1 \vert}

\newcommand\red[1]{\textcolor{red}{#1}}
\newcommand{\translation}[1]{\{\!\vert \, #1 \, \vert\!\}}
\newcommand{\indexedtype}[1]{(\hspace{-.195em}\vert \, #1 \, \vert\hspace{-.195em})}
\newcommand{\indexedtyperev}[1]{ \vert\hspace{-.195em}) \, #1 \, (\hspace{-.195em}\vert}

\newcommand{\lineartypea}{\sigma}
\newcommand{\lineartypeb}{\tau}

\newcommand{\compoundtypea}{\vec{\varphi}}
\newcommand{\compoundtypeb}{\vec{\psi}}
\newcommand{\changeoffibera}{u}
\newcommand{\changeoffiberb}{v}
\newcommand{\ofcourse}[1]{{!_{#1}}}

\newtheorem{theorem}{Theorem}
\newtheorem{lemma}{Lemma}

\begin{document}
\title{Indexed linear logic and higher-order model checking}
\author{Charles Grellois
\institute{ENS Cachan and University Paris Diderot}
\email{grellois@pps.univ-paris-diderot.fr}
 \and Paul-Andr\'e Melli\`es
\institute{CNRS and University Paris Diderot}
\email{mellies@pps.univ-paris-diderot.fr}
}
\def\titlerunning{Indexed linear logic and higher-order model-checking}
\def\authorrunning{C. Grellois and P.-A. Melli\`es}
\maketitle
\begin{abstract}
In recent work, Kobayashi observed
that the acceptance by an alternating tree automaton~$\mathcal{A}$
of an infinite tree~$\mathcal{T}$ generated by a higher-order recursion scheme~$\mathcal{G}$
may be formulated as the typability of the recursion scheme~$\mathcal{G}$
in an appropriate intersection type system associated to the automaton~$\mathcal{A}$.
The purpose of this article is to establish a clean connection between
this line of work and Bucciarelli and Ehrhard's indexed linear logic.
This is achieved in two steps.
First, we recast Kobayashi's result in an equivalent infinitary intersection type system
where intersection is not idempotent anymore.
Then, we show that the resulting type system is a fragment of an infinitary version 
of Bucciarelli and Ehrhard's indexed linear logic.
While this work is very preliminary and does not integrate
key ingredients of higher-order model-checking like priorities,
it reveals an interesting and promising connection between higher-order model checking
and linear logic.
\end{abstract}

\section{Introduction}
%
%
Model-checking is a well-established technique in formal verification,
based on the following model-theoretic procedure.
In order to decide whether a given program~$P$ satisfies a property~$\varphi$ of interest,
one interprets the program~$P$ into an appropriate model and translates 
the property~$\varphi$ into an equivalent automaton~$\mathcal{A}$.
The fact that the program~$P$ satisfies the property~$\varphi$ is then reduced 
to the existence of a successful run of the automaton~$\mathcal{A}$ 
over the interpretation of the program~$P$ in the model, which is decidable.
In the specific case of higher-order model checking, a higher-order program~$P$ 
is modelled as a \emph{higher-order recursion scheme} (HORS)
which generates the tree of all its possible behaviours.
%
%
Recall that given a signature $\Sigma$ and a set of variables $\mathcal{V}$, 
a higher-order recursion scheme \G consists of a set of simply-typed 
non-terminals $\mathcal{N}$,  of an axiom $S \in \mathcal{N}$ of type $o$, and of a set of equations (or rewriting rules) of the form
$$
F \quad = \quad \lambda x_1 \cdots \lambda x_n.\ t \quad \quad \text{(denoted } \mathcal{R}(F) \text{)}
$$
where $t$ is a term of base type $o$ and $F \in \mathcal{N}$ has simple type $\sigma_1 \rightarrow \cdots \rightarrow \sigma_n \rightarrow o$.
One requires moreover that there is exactly one such equation per non-terminal,
and that the simple types of $F$ and of $\mathcal{R}(F)$ coincide.
Every such recursion scheme~$\mathcal{G}$ may be thus seen as a term of the simply typed $\lambda$-calculus with fixpoint operator $Y$.
By definition, the \emph{order} of a recursion scheme is the maximal order of its non-terminal's simple type. 
An example of an order-$2$ scheme over the signature $\Sigma=\{\,a:2 \, , \, b:1 \, , \,  c:0 \, \}$ is
\begin{equation}\label{hors}
\begin{tabular}{rcl}
$S$ & $=$ & $F\ c$\\
$F$ & $=$ & $\lambda x.\, a\ x\ (F\ (b\ x))$\\
\end{tabular}
\end{equation}
The labelled and ranked tree generated by the recursion scheme~$\mathcal{G}$ is called the \emph{value tree} of the scheme. 
It is computed by application of these rules starting from the axiom.
In our illustration, the value tree of the recursion scheme~(\ref{hors}) depicted in Figure \ref{order2tree} 
is obtained as the limit of the rewriting sequence :
$$
S \quad \longrightarrow \quad F\ c \quad \longrightarrow \quad a\ c\ (F\ (b\ c)) \quad \longrightarrow \quad a\ c\ (a\ (b\ c)\ F\ (b\ b\ c)) \quad \longrightarrow \quad \cdots
$$
%
The decidability of monadic second order logic (MSO) over the value trees computed by higher-order recursion schemes 
was established for the first time by Ong \cite{ong} using game semantics.
\begin{figure}[t]
\begin{small}
\centering
\begin{minipage}{.5\textwidth}
  \centering
\begin{tikzpicture}
\Tree [.$a$ $c$ [.$a$ [.$b$ $c$ ] [.$a$ [.$b$ [.$b$ $c$ ] ][.$\vdots$ ] ] ] ] ]
\end{tikzpicture}
  \captionof{figure}{An order-$2$ tree.}
\label{order2tree}
\end{minipage}%
\begin{minipage}{.5\textwidth}
  \centering
 \begin{tikzpicture}
\Tree [.$\ \ \ \ a\ \red{q_0}$ $\ \ \ \ c\ \red{q_1}$ [.$\ \ \ \ a\ \ \red{q_0}$ [.$\ \ \ \ b\ \  \red{q_1}$ $c$ ] [.$\ \ \ \ a\ \ \red{q_0}$ [.$b$ $c$ ] [.$a$ $\vdots$ ] ] [.$\ \ \ \ a\ \ \red{q_2}$ [.$b$ $c$ ] [.$a$ $\vdots$ ] ] ] [.$\ \ \ \ a\ \ \red{q_2}$ [.$b$ $c$ ] [.$a$ $\vdots$  $\vdots$ ] ] ] ]
\end{tikzpicture}
  \captionof{figure}{An alternating run-tree.}
\label{runtree}
\end{minipage}
\end{small}
\end{figure}
Other proofs of the same decidability result were then elaborated, either based
on collapsible pushdown automata~\cite{cpda}, on an intersection type system~\cite{kobayashi-ong}
or on Krivine machines~\cite{salvati-walukiewicz-krivine}.
All these proofs are based on the reduction of the decidability of MSO to the decidability 
of the modal $\mu$-calculus, which is in turn equivalent to the existence of a winning run-tree
of an {alternating} parity automaton~$\mathcal{A}$ on the value tree of the higher-order recursion scheme~$\mathcal{G}$.
Recall that an \emph{alternating} tree automaton proceeds in the same way as a usual top-down tree automaton, 
but with an additional ability: at each step of its exploration, 
the automaton can decide to explore a given subtree of the current node several times, or not at all.
Since each of these explorations may be seen as occurring on a different copy of the same subtree,
everything thus works as if the alternating automaton duplicates the subtree the number of times it explores it.
By way of illustration, keeping the same signature~$\Sigma=\{\,a:2 \, , \, b:1 \, , \,  c:0 \, \}$ as before,
a typical transition will duplicate the rightmost child of a node labelled $a\in\Sigma$ 
and explore the first copy with state $q_0$ and the second copy with state $q_2$:
\begin{equation}
\label{delta}
\delta (q_0, a)\ =\  (1,q_1)  \wedge (2,q_0) \wedge (2,q_2)
\end{equation}
In particular, when applied to the value-tree of the recursion scheme~$\mathcal{G}$ in Figure~\ref{order2tree},
the transition induces a run-tree whose upper nodes are depicted in Figure~\ref{runtree}.
The starting point of this work is the observation of an apparent similarity between this ability 
of alternating automata to duplicate a tree in order to explore it several times 
and the duplication mechanisms associated to the exponential modality of linear logic.
In order to clarify this tentative connection between linear logic and higher-order model checking,
we start from the type-theoretic account of alternating parity automata by Kobayashi and Ong~\cite{kobayashi-ong}.
For simplicity, we prefer to restrict ourselves to Kobayashi's work~\cite{koba09} and do not consider priorities
(or parity conditions) at this stage.
By doing so, we restrict the expressivity of the logic to safety properties.
A treatment of priorities would be possible however, along the lines of our recent observation~\cite{tensorial-logic-with-colours}
that priorities behave in just the same comonadic way as the exponential modality of linear logic.

\subsection*{Plan of the paper}
We start by recalling in \S\ref{section/intersection-types} the intersection type system originally considered by Kobayashi~\cite{koba09}.
The first contribution of the paper is to establish in \S\ref{section/quantitative} a correspondence theorem 
between this type system and a quantitative variant of Kobayashi's type system where intersection is not idempotent anymore.
This preliminary steps leads us to establish in~\S\ref{section/linear-interpretation}
that the quantitative intersection type system is a full fragment of an infinitary version 
of Bucciarelli and Ehrhard's indexed linear logic~\cite{ill1,ill2}.


\section{Intersection types and alternating tree automata}
\label{section/intersection-types}
The type-theoretic account of higher-order model-checking initiated by Kobayashi~\cite{koba09} 
is based on the idea that a transition like (\ref{delta}) may be reflected
by giving to the symbol $a\in\Sigma$, in addition to its simple type $o \rightarrow o \rightarrow o$, 
the refined intersection type $q_1 \rightarrow (q_0 \wedge q_2) \rightarrow q_0$. 
In this approach, every such symbol $a\in\Sigma$ of the signature
will generally have several such refined types, each of them
derived from the transitions~$\delta(q,a)$ associated to each state $q\in Q$ of the underlying
alternating automaton~$\mathcal{A}$.
Note that the existence of an accepting alternating run-tree over the value tree of a recursion scheme~$\mathcal{G}$
involves infinite objects, whose structure can be very complex ---
observe in particular that the order-$2$ tree of Figure~\ref{order2tree} is not regular.
In that respect, the type-theoretic account of the tree automaton~$\mathcal{A}$ has one main benefit:
the refined types defined on the symbols~$a,b,c$ of the signature~$\Sigma$ may be lifted
to every simply-typed $\lambda$-term appearing in the recursion scheme~$\mathcal{G}$.
%
%
By way of illustration, consider again the recursion scheme~(\ref{hors}) and assume
that the alternating automaton~$\mathcal{A}$ has the additional transitions
$$
\delta(q_0,b) \hspace{.5em} =  \hspace{.5em} (1,q_2)
\quad\quad\quad\quad
\delta(q_2,b)  \hspace{.5em} =  \hspace{.5em} (1,q_0)\wedge(1,q_1).
$$
In this situation, the symbol~$b$ of simple type $o\to o$ is given
the refined types $q_2 \rightarrow q_0$ and $(q_0 \wedge q_1) \rightarrow q_2$
and one can thus type the term $\mathcal{R}(F)$ in the following way:
$$
\lambda x.\, a\ x\ (F\ (b\ x))\quad : \quad (q_0 \wedge q_1 \wedge q_2) \rightarrow q_0
$$
under the assumption that the non-terminal $F$ of simple type~$o\to o$
has the refined types $q_0 \rightarrow q_0$ and $q_2 \rightarrow q_2$.
From this lifting property, Kobayashi~\cite{koba09} deduces a decision procedure for the existence 
of a run-tree of the alternating automaton~$\mathcal{A}$ over the value-tree of the recursion scheme~$\mathcal{G}$.
Here, we consider the intersection type system of \cite{koba09} as it is recently rephrased by Ong and Tsukada \cite{ong-tsukada}.
We thus define refined pre-types as follows:
\begin{center}
\begin{tabular}{cccc}
\emph{Refined pre-types} \quad\quad\quad\quad & 
$\sigma, \tau$ & \quad $::=$ \quad & $q \quad | \quad \tau \rightarrow \sigma \quad | \quad \bigwedge_{j\in J} \, \tau_j$
\end{tabular}
\end{center}
Note that in this system refined types have to match the shape of simple types. To ensure this, denoting $\sigma$ a refined type and $\kappa$ a simple type\footnote{In Kobayashi's original article \cite{koba09}, simple types were called \emph{kinds}, and the word "type" was reserved to what we call here refined types.}, we introduce the proper refinement relation $\sigma :: \kappa$, defined by the following rules:
\begin{center}
\begin{tabular}{ccc}
\AxiomC{}
\UnaryInfC{$\vdash \, q::o$}
\DisplayProof
&
\hspace{2cm}
&
\AxiomC{$\vdash \, \tau_j \, :: \, \kappa$ \quad (for all $j\in J$)}
\AxiomC{$\vdash \, \sigma \, :: \, \kappa'$}
\BinaryInfC{$\vdash \, \bigwedge_{j\in J} \tau_j \, \rightarrow \, \sigma :: \,\,  \kappa\rightarrow \kappa'$}
\DisplayProof\\
\end{tabular}
\end{center}
A \emph{refined type} is a refined pre-type which properly refines a simple type. A sequent is of the form
$$
x_1: \tau_1:: \kappa_1 \,  , \,  \dots  \, , \, x_n:\tau_n:: \kappa_n \, \vdash \, M:\sigma::\kappa
$$
where the context
$$
\Gamma \quad = \quad 
x_1: \tau_1:: \kappa_1 \,  , \,  \dots  \, , \, x_n:\tau_n:: \kappa_n
$$
is a sequence of different variables, each of them typed by a refined type and by the simple type it refines. 
The rules of the system are given in Figure \ref{kosys}. 

\begin{figure}[t!]
\begin{center}
\AxiomC{$\vdash \hspace{.4em} \tau_j::\kappa$ \quad (for all $j\in J$)}
\LeftLabel{Axiom \quad \quad}
\RightLabel{$\quad \quad i \in J$}
\UnaryInfC{$\Gamma \, , \, x:\bigwedge_{j\in J}\tau_j::\kappa \hspace{.4em}  \vdash   \hspace{.4em} x:\tau_i::\kappa$}
\DisplayProof
\end{center}
\vspace{-1em}
\begin{center}
\AxiomC{$\Gamma\vdash M:(\bigwedge_{j\in J} \tau_j)\rightarrow \sigma::\kappa\to\kappa'$}
\AxiomC{$\Gamma\vdash N:\tau_j::\kappa$ \quad (for all $j\in J$)}
\LeftLabel{Application \quad \quad}
\BinaryInfC{$\Gamma \vdash MN:\sigma::\kappa'$}
\DisplayProof
\end{center}
\vspace{-1em}
\begin{center}
\AxiomC{$\Gamma, x:\bigwedge_{j\in J} \tau_j ::\kappa  \hspace{.4em} \vdash  \hspace{.4em} M:\sigma::\kappa'$}
\LeftLabel{Lambda \quad \quad}
\UnaryInfC{$\Gamma \vdash \lambda x.M: (\bigwedge_{j\in J} \tau_j)\rightarrow \sigma::\kappa\to\kappa'$}
\DisplayProof
\end{center}
\vspace{-.8em}
%
\caption{Kobayashi's intersection type system.}
\label{kosys}
\end{figure}
The decidability result requires that the set of refinement types~$\sigma$ 
which refine a given simple type~$\kappa$ remains finite.
To that purpose, intersection is required to be \emph{idempotent} in Kobayashi's type system.
This idempotency property may be neatly formulated by requiring that intersections
are stable under \emph{surjective reindexing} $f\,:\,J \, \rightarrow\!\!\!\!\!\!\rightarrow \, I$ in the sense that
$$
\bigwedge_{j \in J} \sigma_{f(j)} \quad = \quad \bigwedge_{i \in I} \sigma_i
$$
for every family $\{ \sigma_i \, | \, i\in I\}$ of refinement types indexed by a finite set~$I$.
Note in particular that the expected equation
$$
\sigma \wedge \sigma \quad = \quad \sigma 
$$
follows from the consideration of the surjective reindexing~$\{1,2\}\to\{1\}$.
At this stage, we make the following observation:
\begin{lemma}
If $\Gamma \vdash t\,:\,\sigma\,::\,\kappa$ in Kobayashi's system  and $t$ $\eta$-expands to $t'$, then $\Gamma \vdash t'\,:\,\sigma\,::\,\kappa$.
\end{lemma}
In other words, typability is preserved by $\eta$-expansion in Kobayashi's type system.
%
%
For that reason, we will only consider $\beta\eta$-long normal form $\lambda$-terms in the sequel.
%
%
%

\section{A quantitative variant of the original type system}
\label{section/quantitative}
%
%
Seen from the point of view of linear logic, Kobayashi's type system appears 
as a variant of natural deduction based on an additive translation of intuitionistic logic.
In order to prepare the forthcoming connection with indexed linear logic, we turn it
into a sequent calculus based this time on a multiplicative translation of intuitionistic logic.
This leads us to the system of Figure \ref{koquant}, where the finite intersections are no longer
required to be stable under surjective reindexing. 
%
Note that we do not even require any associativity or commutativity condition on the intersection:
in particular, the intersections are not even stable under \emph{bijective} reindexing. 

\begin{figure}[t!]
%
%
\begin{center}
\AxiomC{$q \in Q$}
\LeftLabel{Axiom \quad \quad} 
\UnaryInfC{$x:q :: o  \vdash x : q :: o $}
\DisplayProof
\end{center}
\vspace{-1em}
\begin{center}
\AxiomC{$\Gamma , x:\tau :: \kappa  \vdash M : \sigma :: \kappa' $}
\LeftLabel{Left $\bigwedge$ \quad \quad} 
\RightLabel{$\quad \quad i \in \mathbb{N}, \ \tau_i\,=\,\tau$}
\UnaryInfC{$\Gamma , x:\bigwedge_{j \in\{i\}} \tau_j :: \kappa  \vdash M : \sigma :: \kappa' $}
\DisplayProof
\end{center}
\vspace{-1em}
\begin{center}
\AxiomC{$\Gamma_j \vdash M:\tau_j :: \kappa$ \quad (for all $j\in J$)}
\LeftLabel{Right $\bigwedge$ \quad \quad}
\UnaryInfC{$\bigwedge_{j \in J} \Gamma_j \vdash M:\bigwedge_{j\in J} \tau_j :: \kappa $}
\DisplayProof
\end{center}
\vspace{-1em}
\begin{center}
\AxiomC{$\Gamma_1 \vdash N : \bigwedge_{j \in J} \tau_j :: \kappa $}
\AxiomC{$\Gamma_2 , x : \sigma :: \kappa'  \vdash M : \alpha :: \kappa''$}
\LeftLabel{Left $\rightarrow$ \quad \quad}
\RightLabel{\quad \quad $\sigma$ linear}
\BinaryInfC{$\Gamma_1 , \Gamma_2 , f: \left( \bigwedge_{j \in J} \tau_j \right)  \rightarrow \sigma :: \kappa \rightarrow \kappa'  \vdash M[x\,:=\,f\ N] :
\alpha :: \kappa''$}
\DisplayProof
\end{center}
\vspace{-1em}
\begin{center}
\AxiomC{$\Gamma, x: \bigwedge_{j \in J} \tau_j :: \kappa \vdash M : \sigma :: \kappa'$ }
\LeftLabel{Right $\rightarrow$ \quad \quad} 
\UnaryInfC{$\Gamma \vdash \lambda x.M : \left( \bigwedge_{j \in J} \tau_j \right) \rightarrow \sigma :: \kappa \rightarrow \kappa'$}
\DisplayProof
\end{center}
\vspace{-1em}
\begin{center}
\AxiomC{$\Gamma\vdash M:\sigma :: \kappa'$}
\LeftLabel{Weakening \quad \quad}
\RightLabel{\quad \quad ($x \notin \Gamma$)}
\UnaryInfC{$\Gamma, x\,:\,\bigwedge_{j\in \emptyset} \tau_j ::\kappa  \vdash M:\sigma::\kappa'$}
\DisplayProof
\end{center}
\begin{center}
\AxiomC{$\Gamma,\, y\,:\,\bigwedge_{i\in I} \tau_i ::\kappa ,\, z\,:\,\bigwedge_{j\in J} \tau_j ::\kappa  \vdash M:\sigma::\kappa'$}
\LeftLabel{Contraction \quad \quad}
\RightLabel{\quad \quad ($x \notin \Gamma$)}
\UnaryInfC{$\Gamma, x\,:\,\bigwedge_{k\in I \uplus J} \tau_k ::\kappa  \vdash M[y,z := x]:\sigma::\kappa'$}
\DisplayProof
\end{center}
\vspace{-.8em}
\caption{The quantitative intersection type system.}
\label{koquant}
\end{figure}

We say that a variable $x$ occurs \emph{linearly} in a context~$\Gamma$
when the variable~$x$ has refined type $q$ or $\bigwedge_{i \in I} \tau_i \rightarrow \sigma$ in the context~$\Gamma$.
In other words, the variable~$x$ is declared linear when the principal connective of its type is not an intersection.
We say that the variable $x$ occurs \emph{linearly} in a $\lambda$-term $t$ when there exists a unique occurence of the variable $x$ in $t$.

\begin{lemma}
If $\Gamma \vdash t\,:\,\sigma\,::\,\kappa$ and $x$ occurs linearly in $\Gamma$, then $x$ occurs linearly in $t$.
\end{lemma}

An important consequence of the lemma is that the variable~$x$ occurs linearly 
in the $\lambda$-term~$M$ considered in the Left $\rightarrow$ rule.
From this follows that the Left $\rightarrow$ rule which transforms~$M$ 
into the $\lambda$-term~$M[x\,:=\,f\ N]$ introduces exactly one application node~$f\ N$
in the $\lambda$-term~$M$.

Both intersection type systems formulated in Figure~\ref{kosys}
and in Figure~\ref{koquant} are designed to simulate an alternating automaton~$\mathcal{A}$
exploring the value-tree of a higher-order recursion scheme~$\mathcal{A}$.
One main difference comes from the fact that in Kobayashi's system the multiplicity of usage 
of a given state is not tracked, so that a function using its argument twice with refined type $q_0$
in order to answer a request $q_1$ may actually be typed $q_0 \wedge q_2 \rightarrow q_1$. 
This is impossible in the quantitative system, where the Weakening rule only introduces an intersection indexed by the empty family. In order to formalize a precise connection between the type systems, we thus need an appropriate notion of order over qualitative refined types (that is, the idempotent refined types of Kobayashi's system). 
This notion of order is precisely the one of the Scott lattice model of linear logic, see for instance \cite{terui,ehrhard-collapse}:
\begin{itemize}
\item If $\sigma \,::\, o$ and $\tau \,::\, o$, then $\sigma \preccurlyeq \tau$ if and only if $\sigma\ =\ \tau$.
\item Define $\bigwedge_{i \in I} \sigma_i \rightarrow \tau \preccurlyeq \bigwedge_{j \in J} \sigma'_j \rightarrow \tau'$ if and only if both types refine a same simple type $\kappa \rightarrow \kappa'$, $\tau \preccurlyeq \tau'$ and $\forall i \in I\ \ \exists j \in J \quad \sigma'_j \preccurlyeq \sigma_i$.
\end{itemize}
Given a quantitative type $\sigma$, define its \emph{collapse} $\collapse{\sigma}$ as the qualitative type canonically obtained by assuming stability by surjective reindexing. 
This operation is extended to contexts in the standard way. 
We may now give a precise description of the connection between the two type systems.
We will see in the next section that the quantitative type system is in fact designed to reflect the relational semantics of linear logic,
this correspondence theorem may be seen as a type-theoretic transcription of Ehrhard's recent collapse theorem~\cite{ehrhard-collapse}
between the relational semantics of linear logic and its Scott lattice model.
\begin{theorem}\label{theorem/correspondence}
Every derivation tree of one system may be effectively translated in the other either by lifting qualitative types or by collapsing quantitative types:
\begin{itemize}
\item If $\ \Gamma \vdash t\,:\,\sigma\,:\,\kappa$ in the quantitative system, then $\collapse{\Gamma} \vdash t\,:\,\collapse{\sigma}\,::\,\kappa$ in Kobayashi's system.
\item If $\ x_1\,:\,\sigma_1\,::\,\kappa_1, \ldots, x_n\,:\,\sigma_n\,::\,\kappa_n \vdash t\,:\,\tau\,:\,\kappa$ in Kobayashi's system, 
there exists quantitative types $\hat{\sigma_i}$ ($1 \leq i \leq n$) and $\hat{\tau}$ such that
\begin{itemize}
\item $\forall i \in \{1, \ldots, n\}\ \ \hat{\sigma_i}\,::\,\kappa_i$ and $\hat{\tau}\,::\,\kappa$,
\item $\forall i \in \{1, \ldots, n\}\ \ \collapse{\hat{\sigma_i}} \preccurlyeq \sigma_i$ and $\collapse{\hat{\tau}} \preccurlyeq \tau$,
\item $\ x_1\,:\,\hat{\sigma_1}\,::\,\kappa_1, \ldots, x_n\,:\,\hat{\sigma_n}\,::\,\kappa_n \vdash t\,:\,\hat{\tau}\,:\,\kappa$ in the quantitative system.
\end{itemize}
\end{itemize}
\end{theorem}

\section{A linear interpretation of intersection types}
\label{section/linear-interpretation}
In this section, we give an alternative formulation of the fragment of Bucciarelli and Ehrhard's indexed linear logic
necessary to interpret general continuations, and thus higher-order recursion schemes.
The restriction of Bucciarelli and Ehrhard's indexed linear logic \cite{ill1,ill2} to this specific fragment
enables us to annotate every proof of the logic with a simply-typed $\lambda$-term.
This leads us to a formulation of indexed linear logic in the style of de Carvalho~\cite{carvalho}.
%
%
We then explain how to translate Kobayashi's type system into the resulting fragment
of indexed linear logic.

\subsection{Indexed Linear Calculus}
%


Every formula of indexed linear logic is indexed by a countable indexing set~$J$.
As already mentioned, we will focus on the fragment of the logic corresponding
to general continuations, whose formulas are generated by the following grammar:
\begin{center}
\begin{tabular}{cccc}
\emph{Linear pre-formulas} \quad\quad\quad\quad & 
$A, B$ & \quad $::=$ \quad & \quad $\bot_J \quad | \quad S \multimap A$
\\
\emph{Replicable pre-formulas} \quad\quad\quad\quad & 
$S,T$ & \quad $::=$ \quad & \quad $\ofcourse{\changeoffibera}\, A$
\end{tabular}
\end{center}
where~$J$ is any countable set, and where~$\changeoffibera:J\to K$ is any function between the two countable indexing sets~$J$ and~$K$.
Every linear or replicable formula of indexed linear logic is defined as a pre-formula
obtained by a series of application of the rules below.
\begin{center}
\begin{tabular}{ccccc}
\AxiomC{}
\UnaryInfC{$\vdash_{J} \, \, \bot_J$}
\DisplayProof
&
\hspace{1cm}
&
\AxiomC{$\vdash_{J} \, \, S$}
\AxiomC{$\vdash_{J} \, \, A$}
\BinaryInfC{$\vdash_{J} \, \, S\multimap A$}
\DisplayProof
&
\hspace{1cm}
&
\AxiomC{$\vdash_{J} \, \, A$}
\RightLabel{\quad\quad $\changeoffibera:J\to K$}
\UnaryInfC{$\vdash_{K} \, \, \ofcourse{\changeoffibera}\, A$}
\DisplayProof
\\
\end{tabular}
\end{center}
Quite obviously, there exists for every linear formula~$A$
a unique countable indexing set~$J$ such that~$\vdash_{J} \, A$.
This specific countable set~$J$ is called the domain~$dom(A)$ of the formula~$A$.
Now, suppose given a set $Q=\{q_1,\dots, q_n\}$ of elements,
typically representing the states of an alternating automaton~$\mathcal{A}$.
The types of the logic are then defined by the following grammar:
\begin{center}
\begin{tabular}{cccc}
\emph{Linear pre-types} \quad\quad\quad\quad & 
$\lineartypea, \lineartypeb$ & \quad $::=$ \quad & \quad $q \quad | \quad \compoundtypea \multimap \lineartypea$
\\
\emph{Compound pre-types} \quad\quad\quad\quad & 
$\compoundtypea, \compoundtypeb$ & \quad $::=$ \quad & \quad
$[ \, \lineartypea_j \, | \, j\in J \, ]$
\end{tabular}
\end{center}
where~$J$ is any countable set of indices.
A type is then defined as a pre-type which refines
a specific formula of our fragment of indexed linear logic.
The refinement relation is defined by structural induction, and may
be formulated by the following derivation rules.
\begin{center}
\begin{tabular}{ccc}
\AxiomC{$q_j\in Q$ \quad (for all $j\in J$)}
\UnaryInfC{$\vdash_{j\in J} \, \, q_j \, :: \, \bot_{J}$}
\DisplayProof
&
\hspace{2cm}
&
\AxiomC{$\vdash_{j\in J} \, \, \compoundtypea_j\, \, :: \,\, {\ofcourse{\changeoffibera}\, A}$}
\AxiomC{$\vdash_{j\in J} \, \,  \lineartypea_j\, \, :: \,\, B$}
\BinaryInfC{$\vdash_{j\in J} \,  \, \compoundtypea_j\multimap \lineartypea_j \, \, :: \, \, {\ofcourse{\changeoffibera}\, A}\multimap B$}
\DisplayProof\\
\end{tabular}
\end{center}
\begin{center}
\AxiomC{$\vdash_{j\in J} \, \, \lineartypea_j \, :: \, A$}
\RightLabel{\quad $\changeoffibera:J\to K$}
\UnaryInfC{$\vdash_{k\in K} \, \, [\, \lineartypea_{j} \, \, | \, \, \changeoffibera(j)=k \, ] \, \, :: \,\,  \ofcourse{\changeoffibera} \, A$}
\DisplayProof
\end{center}
For conciseness, $\vdash_{i \in I}$ may be abbreviated $\vdash_I$. 
The idea behind this formulation is that a quantitative refinement type $\bigwedge_{k \in K}\ \sigma_k :: \kappa$ 
may be seen as an indexed type
$$
\vdash_{k\in K} \, \, [\, \lineartypea_{j} \, \, | \, \, \changeoffibera(j)=k \, ] \, \, :: \,\,  \ofcourse{\changeoffibera} \, A \quad \quad u\,:\,K \rightarrow 1
$$
where $A$ is a linear formula refining the simple type $\kappa$. Indexing by $u\,:\,K \rightarrow 2$ would give two such intersection types in parallel. The type system of Indexed Linear Calculus (ILC) is described in Figure \ref{ilc}. In the Dereliction rule, the action $u^*$ of a bijection $u\,:\,J \rightarrow K$ is defined by structural induction:
\begin{itemize}
\item $u^*(\bot_K) \ \ = \ \ \bot_J$
\item $u^*(S \multimap A) \ \ = \ \ u^*(S) \multimap u^*(A)$
\item $u^*(!_v\ A) \ \ = \ \ !_{u^{-1} \circ v}\ A$
\end{itemize}
and in the Contraction rule, the domain restriction operation $A\vert_{K}$ where $A$ is of formula of domain $J$ is also defined by structural induction:
\begin{itemize}
\item $\bot_J \, |_{K} := \bot_K$
\item $(\, S \multimap A \, ) \, |_{K} \,  :=  \,  S \, |_{K}  \multimap A \, |_{K}$
\item $(\, {\ofcourse{\changeoffibera}} \, A \, ) \, |_{K} \, =  \, {!_{\changeoffiberb}} \, ( \, A \, |_{\changeoffibera^{-1}(K)} \, )$
where $\changeoffibera:I\to J$ and $\changeoffiberb:\changeoffibera^{-1}(K)\to K$ is equal to the function~$\changeoffibera$
restricted to the subset~$\changeoffibera^{-1}(K)\subseteq I$.
\end{itemize}

\begin{figure}[t!]
\begin{center}
\AxiomC{$q_j\in Q$ \quad (for all $j\in J$)}
\LeftLabel{Axiom\quad \quad} 
\UnaryInfC{$x\,:\,q_j\,::\,\bot_{J} \, \, \vdash_{J} \, \, x\,:\,q_j\,::\,\bot_{J}$}
\DisplayProof
\end{center}
\vspace{-1em}
\begin{center}
\AxiomC{$\Gamma, x\,:\,\phi_j\,::\,!_u\ A \vdash_{J} M\,:\,\sigma_j\,::\,B$}
\LeftLabel{Right $\multimap$ \quad\quad}
\UnaryInfC{$\Gamma \vdash_{J} \lambda x . M \,:\,\phi_j \multimap \sigma_j \,::\, A \multimap B$}
\DisplayProof
\end{center}
\vspace{-1em}
\begin{center}
\AxiomC{$\Gamma_1 \vdash_J M\,:\,\phi_j\,::\,!_u\ A$}
\AxiomC{$\Gamma_2, x\,:\,\sigma_j \,::\,B \vdash_J N\,:\,\tau_j \,::\,C$}
\LeftLabel{Left $\multimap$ \quad\quad}
\BinaryInfC{$\Gamma_1, \Gamma_2, f\,:\,\phi_j \multimap \sigma_j\,::\,!_u\ A \multimap B \vdash_J N[x\,\leftarrow\,f\,M]\,:\,\tau_j \,::\,C$}
\DisplayProof
\end{center}
\vspace{-1em}
\begin{center}
\AxiomC{$\Gamma \hspace{.2em} \vdash_J \hspace{.2em} M\,:\,\sigma_j\,::\, B$}
\LeftLabel{Weakening\quad\quad}
\RightLabel{\quad\quad $\mathbf{0}_J : \emptyset \to J$}
\UnaryInfC{$\Gamma \hspace{.2em},\hspace{.2em} x \,:\, [] \,::\,!_{\, \mathbf{0}_J}\, A\hspace{.2em} \vdash_J\hspace{.2em}M\,:\, \sigma_j\,::\, B$}
\DisplayProof
\end{center}
\vspace{-1em}
\begin{center}
\AxiomC{$\Gamma\hspace{.2em},\hspace{.2em}y \,:\,\phi_j\,::\,!_{u_1} \,  ( \, A \vert_{J_1} \,)
\hspace{.2em},\hspace{.2em} z  \,:\,\psi_j\,::\, !_{u_2} \, ( \, A \vert_{J_2} \, )\hspace{.2em}\vdash_J\hspace{.2em} M \,:\,\sigma_j\,::\, B$}
\LeftLabel{Contraction\quad\quad}
\RightLabel{\quad\quad $u=[u_1,u_2] : J_1+J_2 \to J$}
\UnaryInfC{$\Gamma\hspace{.2em},\hspace{.2em} x  \,:\, \phi_j \uplus \psi_j\,::\,!_u \, A\hspace{.2em}\vdash_J\hspace{.2em}M[y,z\,\leftarrow\,x]\,:\,\sigma_j\,::\, B$}
\DisplayProof
\end{center}
\vspace{-1em}
\begin{center}
\AxiomC{$
u^*\Gamma\hspace{.2em},\hspace{.2em} x\,:\,\sigma_j \,::\,A\hspace{.2em}\vdash_{J}\hspace{.2em}M\,:\,\tau_{u(j)}\,::\,u^* B$}
\LeftLabel{Dereliction\quad\quad}
\RightLabel{\quad\quad$u:J\to K$ bijective}
\UnaryInfC{$\Gamma\hspace{.2em},\hspace{.2em}x\,:\,\sigma_{u^{-1}(k)}\,::\,!_u\, A\hspace{.2em}\vdash_{K}\hspace{.2em} M\,:\,\tau_k\,::\,u^* B$}
\DisplayProof
\end{center}
\vspace{-1em}
\begin{center}
\AxiomC{$\dots,\  \ x_k\,:\,[\sigma_{i_k}\, \vert\, i_k \in I_k,\,u_k(i_k)=j]  \,::\,!_{u_k}\, A_k\,, \ \dots \hspace{.2em}\vdash_J\hspace{.2em}M\,:\,\tau_j\,::\,B$}
\LeftLabel{Promotion\quad\quad}
\RightLabel{$\quad\quad v:J\to L$}
\UnaryInfC{$\dots,\ \ x_k\,:\,[\sigma_{i_k}\, \vert\, i_k \in I_k,\,v(u_k(i_k))=l]  \,::\,!_{v \circ u_k}\, A_k\,,\ \dots \hspace{.2em}\vdash_L\hspace{.2em}M\,:\,[\tau_j\,\vert\,v(j)=l]\,::\,!_{v}\ B$}
\DisplayProof
\vspace{-.4em}
\caption{Indexed Linear Calculus.}
\label{ilc}
\end{center}
\end{figure}

Note that these operations were only defined on formulas. Their action on types is the expected one: for the domain restriction operation, a compound type $(\phi_j)_{j \in J}$ is restricted to $(\phi_j)_{j \in K}$, and for the other operation, the bijection naturally acts by reindexing over compound types.

\subsection{Interpreting quantitative intersection types in ILC}

Now translating the quantitative intersection types into indexed types is essentially immediate: given a quantitative type $\sigma$, define the corresponding indexed type $\indexedtype{\sigma}$ inductively as follows :
\begin{itemize}
\item $\indexedtype{\bigwedge_{i \in I}\ \sigma_i}\ =\ [\indexedtype{\sigma_i}\,\vert\, i \in I]\quad \quad$ (seen as a $\{\star \}$-indexed formula)
\item $\indexedtype{\sigma \rightarrow \tau}\ =\ \indexedtype{\sigma} \multimap \indexedtype{\tau}$
\item $\indexedtype{q}\ =\ q$
\end{itemize}
This operation can be inverted for $\{\star \}$-indexed formulas in the expected way.
Given a $\{\star\}$-indexed type $\sigma$, the corresponding quantitative intersection type is denoted $\indexedtyperev{\sigma}$.

 Lifting simple types to formulas requires to index formulas properly. For example, $q_1 \wedge q_2 \rightarrow q_0$ refines the simple type $o \rightarrow o$. This lifts to the linear type $[q_1, q_2] \multimap q_0$ refining the formula $!_u\ \bot_2 \multimap \bot_1$ where $u$ is the unique function $\{1,2\} \rightarrow \{1\}$. Given a formula of indexed linear logic $A$, we define inductively its corresponding simple type $\kappa(A)$ as follows:
\begin{itemize}
\item $\kappa(!_u\ A) \ =\ \kappa(A)$
\item $\kappa(S \multimap A)\ = \ \kappa(S) \rightarrow \kappa(A)$
\item $\kappa(\bot_J)\ =\ o$
\end{itemize}

Given a context $\Gamma\,=\,x_1 \,:\,\sigma_1\,::\,A_1\,,\,\ldots\,,\, x_n \,:\,\sigma_n\,::\,A_n$ of indexed linear calculus, define its associated quantitative intersection context $\translation{\Gamma}\ =\ x_1\,::\,\indexedtyperev{\sigma_1}\,::\,\kappa(A_1)\,,\,\ldots\,,\,x_n\,::\,\indexedtyperev{\sigma_n}\,::\,\kappa(A_n)$.\\

\begin{theorem}\label{theorem/comparison}
\begin{itemize}
\item If the sequent $\Gamma\ \vdash_I\ M\,:\,\sigma\,::\,A$, where $I$ is a singleton, is provable in the indexed linear calculus, then $\translation{\Gamma}\ \vdash\ M\,:\,\indexedtyperev{\sigma}\,::\,\kappa(A)$ is provable in the quantitative intersection system.
\item  If the sequent $\Gamma\ \vdash\ M\,:\,\sigma\,::\,\kappa$ is provable in the quantitative intersection system, there exists a context $\Gamma'$ of the indexed linear calculus and a formula $A$ of indexed linear logic such that $\translation{\Gamma'}\ =\ \Gamma$ and that $\Gamma'\ \vdash_{\{\star\}}\ M\,:\,\indexedtype{\sigma}\,::\,A$ is provable in the indexed linear calculus.
\end{itemize}
\end{theorem}

Recall that indexation is a way to parallelize proofs with the same underlying tree, and only differing by their types labelling. This is the reason why the connection only makes sense for indexation families isomorphic to $\{\star \}$.

Remark now that the indexed linear calculus contains a lot of redundant information. Types may be computed from proof-trees, by picking the state information at the right axiom rule, and terms can be recovered from the rules of a proof-tree, since indexation ensures uniformity of terms when applying them -- contrary to what occurs in the resource lambda-calculus. In fact, this indexed linear calculus captures precisely the fragment of the relational model corresponding to simply-typed $\lambda$-terms. Call \emph{indexed tensorial logic} the system obtained by removing terms and types in the indexed linear calculus. Then derivation trees of this logic are in bijection with ILC derivation trees.

%

\section{Related works}\label{section/related-works}
The starting point of this work is the observation by Kobayashi \cite{koba09} that MSO model-checking over trees generated by higher-order recursion schemes can be performed by typing the scheme in an appropriate intersection type system.

Another inspiration was the recent development by Terui~\cite{terui} of a semantic and type-theoretic approach 
based on linear logic, intersection types and automata theory
in order to characterize the complexity of evaluation to the booleans
in the simply-typed $\lambda$-calculus -- relating on the qualitative model of Scott domains for linear logic. A quantitative account of intersection type in the relational model of linear logic was given by de Carvalho~\cite{carvalho}, also with the goal of studying complexity issues.

After Ong's seminal proof \cite{ong} of the decidability of higher-order model-checking, several other were given. One of them, by Kobayashi and Ong \cite{kobayashi-ong}, extends Kobayashi's type system to capture all MSO. Salvati and Walukiewicz are currently developing
a semantic approach to higher-order model checking, based on the interpretation
of the Krivine environment machine in finite models of the $\lambda$-calculus 
with fixpoint operators, in order to obtain a semantic proof of this decidability result~\cite{salvati,salvati2}.

Finally, Tsukada and Ong \cite{ong-tsukada} introduced two-level game semantics to provide a model of Kobayashi's type system.

\section{Conclusion and future work}
The main purpose and contribution of the paper is to stress the tight and somewhat surprising connection
between a series of recent advances in linear logic
and the current type-theoretic approach to higher-order model-checking.
In particular, 
the lucid reader will recognize that all the results stated in the present paper
are essentially known to one community or to the other.
However, besides the useful confrontation of two related lines of research, we believe
that a tangible technical contribution of the paper is a careful proof of Theorem~\ref{theorem/comparison}.
The result is somewhat folklore and appears implicitly in the works by Bucciarelli and Ehrhard~\cite{ill1,ill2}
and by de Carvalho~\cite{carvalho} but it was never proved (nor even formally stated) as far as the authors know.
Another point: besides the connection between indexed linear logic and higher-order model checking,
one main message of the paper conveyed in Theorem~\ref{theorem/correspondence}
is that the decidability results obtained in the field of higher-order model checking
are to a large part regulated by the collapse theorem recently established by Ehrhard~\cite{ehrhard-collapse,ehrhard-collapse-2}.
We hope that the bridge with linear logic will clarify the constructions of higher-order model checking 
and reveal the deep and beautiful semantic ideas underlying it.
%
%
%

\bibliographystyle{eptcs}
\bibliography{itrs}

\end{document}